\documentclass[a4paper,hyper]{JHEP3}

\usepackage[utf8x]{inputenc}
\usepackage{graphicx}
\usepackage{latexsym}
\usepackage{amsmath}

\usepackage{pdfsync}

\newcommand{\Tr}{\textrm{Tr}\,}

\newcommand{\ev}[1]{\langle #1 \rangle}

\newcommand{\nr}[1]{(\ref{#1})}
\newcommand{\csw}{c_{\rm{sw}}}
\newcommand{\eq}[1]{Eq.~(\ref{#1})}

\newcommand{\be}{\begin{equation}}
\newcommand{\ee}{\end{equation}}
\newcommand{\bea}{\begin{eqnarray}} 
\newcommand{\eea}{\end{eqnarray}}

\newcommand{\bmp}{\noindent\begin{minipage}{16cm}}
\newcommand{\emp}{\end{minipage}\vskip 7mm} 
\def\lsim{\mathrel{\raise.3ex\hbox{$<$\kern-.75em\lower1ex\hbox{$\sim$}}}}
\def\gsim{\mathrel{\raise.3ex\hbox{$>$\kern-.75em\lower1ex\hbox{$\sim$}}}}

\newcommand{\intron}[1]{}

\title{Determining the conformal window:
  SU(2) gauge theory with $N_f = 4$, $6$ and $10$ fermion flavours}
\author{Tuomas Karavirta\footnote{tuomas.karavirta@jyu.fi}\\
Department of Physics, P.O.Box 35 (YFL), 
        \\ FI-40014 University of Jyv\"askyl\"a, Finland, 
        \\ and 
  	    \\ Helsinki Institute of Physics, P.O.~Box 64, 
  	    \\ FI-00014 University of Helsinki, Finland.}
\author{Jarno Rantaharju\footnote{jarno.rantaharju@helsinki.fi}\\
 Department of Physics and Helsinki Institute of Physics,\\
 P.O.Box 64, FI-00014 University of Helsinki, Finland}
\author{Kari Rummukainen\footnote{kari.rummukainen@helsinki.fi}\\
 Department of Physics and Helsinki Institute of Physics,\\
 P.O.Box 64, FI-00014 University of Helsinki, Finland}
\author{Kimmo Tuominen\footnote{kimmo.i.tuominen@jyu.fi}\\
Department of Physics, P.O.Box 35 (YFL), 
        \\ FI-40014 University of Jyv\"askyl\"a, Finland, 
        \\ and 
  	    \\ Helsinki Institute of Physics, P.O.~Box 64, 
  	    \\ FI-00014 University of Helsinki, Finland.}
\abstract {%
  We study the evolution of the coupling in SU(2) gauge field theory with
  $N_f=4$, $6$ and $10$ fundamental fermion flavours on the lattice.
  These values are chosen close to the expected edges of the
  conformal window, where the theory possesses an infrared fixed point.
  We use improved Wilson-clover action, and measure the coupling in the
  Schr\"odinger functional scheme.  At four flavours we observe that the
  coupling grows towards the infrared, implying QCD-like behaviour,
  whereas at ten flavours the results are compatible with a Banks-Zaks type
  infrared fixed point. The six flavour case remains inconclusive:
  the evolution of the coupling is seen to become slower at the infrared, 
  but the accuracy of the results falls short from fully resolving the fate
  of the coupling. We also measure the mass anomalous dimension 
  for the $N_f=6$ case.
}


\begin{document}


\section{Introduction}

There has recently been interest in studying quantum field theories
with a nontrivial infrared fixed point (IRFP), both in continuum and on
the lattice. Under the renormalization group evolution, the coupling 
of these theories shows asymptotic
freedom at small distances, analogously to QCD, but flows to a fixed
point at large distances where the theory looks conformal. Such theories have
applications in beyond Standard Model model building. These include
unparticles, i.e. an infrared conformal sector coupled weakly to the
Standard Model \cite{Georgi:2007ek,Georgi:2007si,Cheung:2007zza,Sannino:2008nv}, and
extended technicolor scenarios, that explain the masses of the
Standard Model gauge bosons and fermions via strong coupling gauge
theory dynamics \cite{TC,Eichten:1979ah,Hill:2002ap,Sannino:2008ha}.

\begin{figure}
\centering
\includegraphics[scale=0.6]{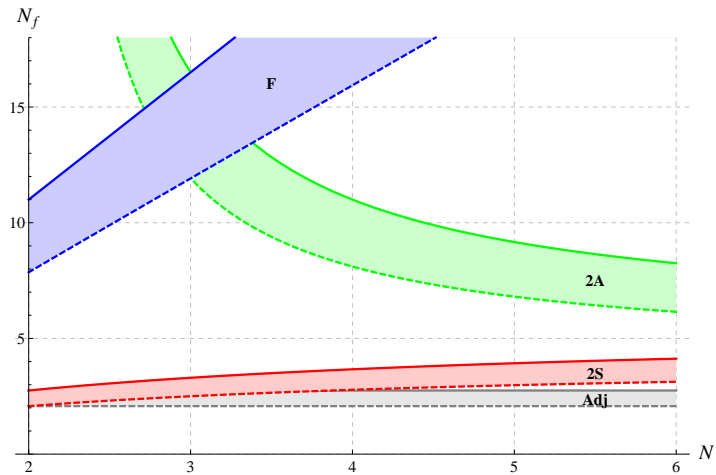}
\caption{The phase diagram of SU($N$) gauge theory as a function of the number of colours, flavours and fermion representations (F = Fundamental,
2A = 2-index antisymmetric, 2S = 2-index symmetric, Adj = Adjoint).
The shaded bands indicate the estimated conformal windows.}
\label{phasediagram}
\end{figure}

In addition to direct applications to particle phenomenology, the phase diagrams of gauge theories, as a function of the number of colours, $N$, flavours $N_f$ and fermion representations, are interesting from the purely theoretical viewpoint of understanding the nonperturbative gauge theory dynamics from first principles.  In figure \ref{phasediagram} we show a sketch of such phase diagram for SU($N$) gauge theory \cite{Sannino:2004qp}. In addition to the fundamental representation, the figure shows the phase structure in the cases of two-index (anti)symmetric and adjoint representations.  The shaded regions in the figure depict the conformal windows for each of these fermion representations; below the conformal window the theory is in the chiral symmetry breaking and confining phase, while above the conformal window the theory is in the non-Abelian QED-like Coulomb phase. The upper boundary in each case corresponds to the loss of asymptotic freedom, i.e. to the value of $N_{f}$ where the one-loop
coefficient $\beta_0=11/3 N-4/3 N_{f} T(R)$ of the $\beta$-function vanishes. The group theory factor $T(R)$ is defined for each representation as
\begin{equation}
\textrm{Tr}(T^aT^b)=T(R)\delta^{ab}.
\end{equation}
For fundamental (F), two-index symmetric (2S), two-index antisymmetric (2AS) and adjoint (A), the concrete values are, respectively, $T(F)=1/2$, $T(2S)=(N -
2)/2$, $T(2AS)=(N + 2)/2$ and $T(A)=N$.

For values of $N$ and $N_f$ inside the conformal window, the theory is
expected to have a non-trivial IRFP. Near the upper boundary one
expects the value of the coupling at the fixed point, $\alpha^\ast$,
to be small and perturbation theory to be applicable
\cite{Banks:1981nn}.  However, as one moves deeper into the conformal
window, by e.g.  lowering $N_f$ at fixed $N$, the fixed point coupling
grows, and nonperturbative methods are required for the analysis. 
In figure \ref{phasediagram} we have used the traditional estimate of evaluating
the critical coupling for chiral symmetry breaking in the ladder
approximation and setting it equal to the fixed point value of the
two-loop coupling. In other words, at the lower boundary of the conformal
window the fixed point coupling becomes super critical with respect to
the critical coupling for the onset of chiral symmetry breaking. 

Explicitly, at two loops we have 
\bea
  \beta(g)\equiv\mu\frac{dg}{d\mu} &=&-\frac{\beta_0}{16\pi^2}g^3 
  - \frac{\beta_1}{(16\pi^2)^2}g^5, \nonumber\\
  \beta_0 &=& \frac{11}{3}N-\frac{4}{3}T(R)N_f,\nonumber \\
  \beta_1 &=& \frac{34}{3}N^2-\frac{20}{3}N\,T(R) N_f-4 C_2(R)T(R) N_f,
\eea 
which implies that IRFP is at 
\be 
  \alpha^\ast=-\frac{\beta_0}{\beta_1}(4\pi).
\ee 
Above, $C_2(R)$ is the quadratic Casimir operator in
representation $R$ and, for the representations we consider, it
assumes values $C_2(F)=(N^2-1)/(2N)$, $C_2(2(A)S)=(N\pm 1)(N\mp 2)/N$
and $C_2(A)=N$.  The estimated critical coupling  on the other hand is
$\alpha_c=\pi/(3C_2(R))$ \cite{Appelquist:1986an}. Solving 
$ \alpha^\ast=\alpha_c$, one obtains
the lower boundary of the conformal window, denoted by dashed lines in figure \ref{phasediagram}. The phase
diagram in the figure \ref{phasediagram} is therefore a conjecture
which must be checked by nonperturbative analysis. Since currently the
lattice simulations provide the only robust nonperturbative method for
non-supersymmetric four dimensional gauge theories, such a check
provides an interesting challenge for the lattice community.

In this work we study SU(2) gauge field theory with $N_f =4, 6$ and $10$ massless flavours of fermions in the fundamental representation on the lattice using $O(a)$ improved Wilson-clover fermions.  On the basis of two-loop beta-function six and ten flavour SU(2) theories may have a fixed point and be within the conformal window, while the four flavour theory is expected to confine.  The model with ten fermions is chosen because it is close to the upper edge of the conformal window ($N_f=11$, where asymptotic freedom is lost) and is expected to have a fixed point at a rather small coupling. Thus, in this case we can compare the results from lattice simulations with those obtained from perturbative computations.  The universal 2-loop $\beta$-function has a zero at $g^2 \approx 2.90$, and at 3 or 4 loops in the MS-scheme it is at $g^2 \approx 2.47$ or $2.52$, respectively \cite{vanRitbergen:1997va}.  While the MS-scheme results cannot be directly compared with the Schr\"odinger functional results, the
convergence indicates that the perturbative result should be reasonably accurate.

In the model with six fermions, however, the fixed point determined by the two loop beta function is at much larger coupling, where we cannot expect the perturbation theory to hold.  Indeed, the location of the fixed point varies significantly at different loop levels: using the MS-scheme $\beta$-function at 2, 3 and 4 loops, the location is $g_{\rm FP}^2 \approx 140$, $21$ and $30$, respectively. 
Such a large value of the fixed point coupling may be super critical with respect to the critical coupling for the onset of spontaneous chiral symmetry breaking; this is explicitly illustrated in figure \ref{phasediagram}, where the two colour and six flavour theory is already outside the conformal window. However, due to the nonperturbative nature of the lower boundary of the conformal window, one cannot exclude the possibility that the six flavour theory is very close or even within the conformal window.    In addition, if we treat $N_f$ as a continuously variable parameter, the fixed point vanishes from the 4-loop $\beta$-function when $N_f$ is lowered only slightly from 6 to 5.945. In the proximity of this value the 4-loop $\beta$-function obtains a shape typical of walking coupling, i.e. the $\beta$-function has a local maximum with a small negative value.  Clearly, these results imply that the six flavour case can be resolved only with non-perturbative calculations.

On the other hand, the four flavour case is expected to be QCD-like, with confinement and chiral symmetry breaking.  It is included here for comparison with the larger $N_f$ cases.

On the lattice, SU(2) gauge field theory with $N_f > 2$ fundamental representation fermions has been recently studied in Refs.~\cite{Bursa:2010xr,Ohki:2010sr,Voronov}.%
\footnote{%
  In related work, the existence of the infrared fixed point in
  SU(2) gauge theory with two adjoint representation fermions (which
  is of interest for technicolor model building) has 
  been studied in 
  \cite{Catterall:2007yx,Hietanen:2008mr,
    DelDebbio:2008zf,Catterall:2008qk,Hietanen:2009az,
    Bursa:2009we,DelDebbio:2009fd,DelDebbio:2010hx,
    DelDebbio:2010hu,Bursa:2011ru,DeGrand:2011qd},
  and SU(3) gauge with various fermion representations and 
  numbers of flavours in
  \cite{Damgaard:1997ut,Appelquist:2007hu,Appelquist:2009ty,
    Fodor:2009wk,Deuzeman:2008sc,Deuzeman:2009mh,Itou:2010we,
    Jin:2010vm,Hayakawa:2010yn,Hasenfratz:2010fi,Hasenfratz:2009ea,
    Shamir:2008pb,DeGrand:2008kx,DeGrand:2010na,
    Fodor:2009ar,Kogut:2010cz}.%
}
The study by Bursa {\em et al.} \cite{Bursa:2010xr} of SU(2) with six fundamental flavours suggested a possibility of an IRFP at much smaller coupling than expected from perturbation theory. However, this work used unimproved Wilson fermions which can be expected to be subject to large discretization errors.  Indeed, as described below, the leading order perturbative analysis reveals that the running coupling measurement (step scaling using Schrödinger functional) using unimproved Wilson action can have finite cutoff effects of order 30-60\%, whereas the improved action reduces these to a few percent level using lattices presently within computational reach.   In our simulations using $O(a)$ improved Wilson-clover fermions we indeed observe large deviations from the unimproved results.

We measure the running coupling using the Schr\"odinger functional
method in theories with four, six and ten flavours, and the mass anomalous dimension in the phenomenologically most interesting six flavour case.  Unfortunately, in the six flavour case we are not able to fully resolve whether the fixed point exists, but the possible locations of the fixed point moves to significantly larger coupling ($g^2 \gsim 10$) than the unimproved lattice action results indicate.

This paper is structured as follows:
In section \ref{model}  we introduce the model and describe the
Schr\"odinger functional method, and present the perturbative 
step scaling function results.  The lattice simulations and results are presented in section \ref{measurements} and in section \ref{conclusions} we conclude.

\section{The model and the Schr\"odinger functional method}
\label{model}

The theory is defined by the action
\begin{equation}
  S=S_G + S_F,
  \label{eq:action}
\end{equation}
where $S_G$ is the standard Wilson single plaquette gauge action for the
SU(2) Yang Mills theory
\begin{equation}
   S_G = \beta_L \sum_{x;\mu<\nu} 
   \left (1 - \frac12 \Tr [U_\mu(x) U_\nu(x+a\hat\mu) 
  U^\dagger_\mu(x+a\hat\nu) U^\dagger_\nu(x) ] \right),
\end{equation}
with $\beta_L=4/g_0^2$. The part
$S_F$ is the clover improved Wilson fermion action
\begin{equation}
  S_F
  = a^4\sum_{\alpha=1}^{N_f} \sum_x \left [
  \bar{\psi}_\alpha(x) ( i D + m_0 )
  \psi_\alpha(x)
   + a \csw \bar\psi_\alpha(x)\frac{i}{4}\sigma_{\mu\nu}
  F_{\mu\nu}(x)\psi_\alpha(x) \right ],
\end{equation}
where $D$ is the standard Wilson-Dirac operator
\begin{equation}
  D=\frac12 [\gamma_\mu(\nabla_\mu^* + \nabla_\mu ) - 
  a\nabla_\mu^*\nabla_\mu],
\end{equation}
with the gauge covariant forward and backward lattice derivatives
\begin{equation}
  \nabla_\mu\psi(x) =1/a[U_\mu(x)\psi(x+a\hat\mu) - \psi(x)],  ~~~~
  \nabla_\mu^*\psi(x) =1/a[\psi(x) -
 U^\dagger_\mu(x-a\hat\mu)\psi(x-a\hat\mu)].
\end{equation}
The clover improvement term contains the usual symmetrized field strength tensor.  We set the improvement coefficient $\csw$ to the $N_f$-independent perturbative value \cite{Luscher:1996vw}
\begin{align}
  \csw = 1 + 0.1551(1) g_0^2 + O(g_0^4).
  \label{pertcsw}
\end{align}
While $\csw$ can be determined nonperturbatively \cite{Luscher:1996ug}, in our tests we observed that at strong lattice coupling the nonperturbatively determined $\csw$ is close to the perturbative one at $N_f=6$ and $10$.  The situation is very different at $N_f=2$ \cite{Karavirta:2010ym,Karavirta:2011mv}, where $\csw$ diverges as $g_0^2$ increases.  Thus, while the perturbative result (\ref{pertcsw}) does not give full cancellation of the $O(a)$ discretization effects, it is a concrete recipe which we expect to eliminate most of the $O(a)$ effects.  
We also include the perturbative improvement at the Schr\"odinger functional 
boundaries as described in \cite{Karavirta:2010ef,Karavirta:2011mv}.

We measure the running coupling using the Schr\"odinger functional method
\cite{Luscher:1992an,Luscher:1992ny,Luscher:1993gh,DellaMorte:2004bc}. 
The coupling is defined as a response to the change
of the background field and the scale is set by the finite size of the 
lattice. We consider a lattice of volume $V=L^4=(Na)^4$. The spatial links at the 
$t=0$ and $t=L$ boundaries are fixed to \cite{Luscher:1992ny}
\begin{align}
U_\mu(\bar x,t=0) &= e^{-i\eta \sigma_3 a/L} \\
U_\mu(\bar x,t=L) &= e^{-i(\pi - \eta) \sigma_3 a/L}
\end{align}
with $\sigma_3$ the third Pauli matrix.
The spatial boundary conditions are periodic for the gauge field.
The fermion fields are set to vanish at the $t=0$ and $t=L$ boundaries and have twisted 
periodic boundary conditions to spatial directions: 
$\psi(x + L\hat i) = \exp(i\pi/5)\psi(x)$.

At the classical level the boundary conditions generate a constant chromoelectric 
field and the derivative of the action with respect to $\eta$ can be easily calculated:
\begin{align}
 \frac{\partial S^{\textrm{cl.}}}{\partial \eta} = \frac{k}{g^2_0},
\end{align}
where $k$ is a function of $N=L/a$ and $\eta$ \cite{Luscher:1992ny}. At the full quantum 
level the coupling is defined by
\begin{align}
  \ev{ \frac{\partial S}{\partial \eta} } = \frac{k}{g^2}.
\end{align}

To quantify the running of the coupling we use the step scaling function $\Sigma(u,s,L/a)$ introduced in \cite{Luscher:1992an}. It characterizes the change of the measured coupling when the linear size of the system is changed from $L$ to $sL$ while keeping the bare coupling $g_0^2$ (and hence lattice
spacing) constant:
\begin{align}
 &\Sigma(u,s,L/a) = \left. g^2(g_0^2,sL/a) \right |_{g^2(g_0^2,L/a)=u}
 \label{stepscaling}\\
 &\sigma(u,s) = \lim_{a/L\rightarrow 0} \Sigma(u,s,L/a)
\end{align}
In this work we choose $s=2$. To obtain the continuum limit $\sigma(u,s)$ we calculate $\Sigma(u,s,L/a)$ at $L/a=6$ and $8$.  Since we expect the discretization errors to be (mostly) removed to the first order in the
improved action, we use quadratic extrapolation to find the limit $a\rightarrow 0$.

The step scaling function is related to the $\beta$-function by
\begin{align}
  -2\ln(s) = \int_u^{\sigma(u,s)} \frac{dx}{\sqrt x \beta(\sqrt x)}.
\end{align}
Close to the fixed point, where the running is slow and $|\beta|$ small, we can approximate the $\beta$-function by
\begin{align}
  \beta(g) \approx \beta^*(g) =
 \frac{g}{2\ln(s)} \left ( 1 - \frac{\sigma(g^2,s)}{g^2} \right ). \label{eq:beta*}
\end{align}
The estimating function $\beta^*(g)$ is exact at a fixed point but deviates
from the actual $\beta$-function as $|g-g^\ast |$ becomes large.

We also measure the mass anomalous dimension $\gamma=d\ln m_q/d\ln\mu$ of the 
theory with 6 fermion flavours 
using the pseudoscalar density renormalization constant.  The mass anomalous dimension is of interest for this case, because the possible infrared fixed point is at non-perturbative value of the coupling.  It is defined on the lattice as \cite{DellaMorte:2005kg}
\begin{align}
Z_P(L) = \frac{\sqrt{3} f_1}{f_P(L/2)},
\end{align}
where $f_1$ and $f_P$ are correlation functions of the pseudoscalar density
\begin{align}
f_1 &= \frac{-1}{12 L^6} \int d^3u d^3v d^3y d^3z 
  \ev{\bar \zeta'(u)\gamma_5\lambda^a\zeta'(v)\bar\zeta(y)\gamma_5\lambda^a\zeta(z)},\\
f_P(x_0) &= \frac{-1}{12 L^6} \int d^3y d^3z 
  \ev{\bar \psi(x_0)\gamma_5\lambda^a\psi(x_0)\bar\zeta(y)\gamma_5\lambda^a\zeta(z)}.
\end{align}
For these measurements the boundary matrices at $t=0$ and $t=L$ are set to unity; thus, separate simulations are needed.  
The mass step scaling function is then defined as in \cite{Capitani:1998mq}:
\begin{align}
 \Sigma_P(u,s,L/a) &= 
    \left. \frac {Z_P(g_0,sL/a)}{Z_P(g_0,L/a)} \right |_{g^2(g_0,L/a)=u}
    \label{Sigmap}\\
 \sigma_P(u,s) &= \lim_{a/L\rightarrow 0} \Sigma_P(u,s,L/a),
\end{align}
and we choose again $s=2$. We find the continuum step scaling function $\sigma_P$ by
measuring $\Sigma_P$ at $L/a=6,8$ and $10$ and doing a quadratic extrapolation.

The mass step scaling function is related to the anomalous dimension by
(see \cite{DellaMorte:2005kg})
\begin{equation}
 \sigma_P(u,s) = \left ( \frac{u}{\sigma(u,s)} \right ) ^{d_0/(2b_0)}
   \exp \left [-\int_{\sqrt u}^{\sqrt{\sigma(u,s)}} dx
   \left ( \frac{\gamma(x)}{\beta(x)} - \frac{d_0}{b_0 x} \right )   \right ],
   \label{massstep}
\end{equation}
where $b_0=\beta_0/(16\pi^2)$ in terms of the one-loop coefficient $\beta_0=22/3-2 N_f/3$ of the beta function and 
$d_0=3 C_2(F) g^2/(8\pi^2)=9/(32\pi^2)$ is the corresponding one-loop coefficient for the anomalous dimension, 
$\gamma=-d_0 g^2$.
Close to the fixed point (\ref{massstep}) simplifies considerably.
Denoting the function estimating the anomalous dimension $\gamma(u)$ by
$\gamma^*(u)$, we have
\begin{align}
\log \sigma_P(g^2,s) &= -\gamma^*(g^2)\int_\mu^{s\mu} \frac{d\mu'}{\mu'} = -\gamma^*(g^2)\log s , \\
 &\Rightarrow \gamma^*(g^2) = -\frac{\log \sigma_P(g^2,s)}{\log s }.
 \label{gammastar}
\end{align}
The estimator $\gamma^*(g^2)$ is exact only at a fixed
point where $\beta(g^2)$ vanishes and deviates from the actual anomalous dimension when $\beta$ is large.

The Schr\"odinger functional boundary conditions enable one to run at vanishing quark mass.
The Wilson fermion action breaks the chiral symmetry explicitly and 
allows additive renormalization of the quark mass. We therefore find the quark mass from
the PCAC relation
\begin{align}
aM(x_0) = \frac14 \frac{(\partial_0^* + \partial_0) f_A(x_0)}{f_P(x_0)} +
 c_A \, \frac a 2 \frac{\partial_0^* \partial_0 f_P(x_0)}{f_P(x_0)}.
\end{align}
We have used here the improved axial current
\begin{align}
f_A^I &= f_A + a \, c_A \, \frac{1}{2} (\partial_\mu^* + \partial_\mu) f_P, \\
f_A(x_0) &= \frac{-1}{12 L^6} \int d^3y d^3z 
  \ev{\bar \psi(x_0)\gamma_0\gamma_5\lambda^a\psi(x_0)
  \bar\zeta(y)\gamma_0\gamma_5\lambda^a\zeta(z)}.
\end{align}
In this work use the perturbative expansion for the improvement 
coefficient $c_A$
\cite{Luscher:1996vw}:
\begin{equation}
  c_A = -0.00567(1) C_2(F) g_0^2 + \mathcal{O}(g_0^4),
  ~~~~ C_2(F) = 3/4.
\end{equation}

The Schr\"odinger functional boundary conditions remove the zero modes that would 
normally make it impossible to run simulations at zero mass. We define $\kappa_c$ as 
the value of the parameter $\kappa$ where the mass $aM(L/2)$ vanishes. To find 
$\kappa_c$ we measure the mass at 3 to 7 values of $\kappa$ on lattices of size 
$L/a=16$ and interpolate to find the point where the mass becomes zero. We then use 
the same value of $\kappa_c$ on all lattice sizes. The values of $\kappa_c$ used in the
simulations are given in table \ref{table:kappa}.  In practice we achieve $|aM| < 0.01$.

\begin{table}
\centering
\begin{tabular}{|l|l|l| c |l|l|l| c |l|l|l|}
\cline{1-3} \cline{5-7} \cline{9-11}
\multicolumn{3}{|c|}{$N_f=4$} & & \multicolumn{3}{|c|}{$N_f=6$} & & \multicolumn{3}{|c|}{$N_f=10$}  \\
\cline{1-3} \cline{5-7} \cline{9-11}
 $\beta_L$ & $\kappa_c$  & $N_{traj}$ & & $\beta_L$ & $\kappa_c$ & $N_{traj}$ & & $\beta_L$ & $\kappa_c$ & $N_{traj}$     \\ \cline{1-3} \cline{5-7} \cline{9-11}
 1.8 & 0.14162  & 108755 & & 1.39 & 0.144351377 & 215119 & & 1   & 0.14199 & 216356  \\ \cline{1-3} \cline{5-7} \cline{9-11}
 1.9 & 0.139914 & 89064  & & 1.4  & 0.139914    & 221273 & & 1.3 & 0.13922 & 232495  \\ \cline{1-3} \cline{5-7} \cline{9-11} 
 2   & 0.138638 & 55031  & & 1.44 & 0.14350583  & 209124 & & 1.5 & 0.13762 & 100476  \\ \cline{1-3} \cline{5-7} \cline{9-11}
 2.2 & 0.136636 & 13294  & & 1.5  & 0.142446    & 233273 & & 1.7 & 0.13627 & 99792   \\ \cline{1-3} \cline{5-7} \cline{9-11}
 2.4 & 0.135205 & 23359  & & 1.8  & 0.1385229   & 25886  & & 2   & 0.13466 & 213290  \\ \cline{1-3} \cline{5-7} \cline{9-11}
 3   & 0.132548 & 64035  & & 2    & 0.1367336   & 54710  & & 3   & 0.13146 & 99918   \\ \cline{1-3} \cline{5-7} \cline{9-11}
 4   & 0.130326 & 58879  & & 2.4  & 0.1342875   & 33818  & & 4   & 0.12981 & 100328  \\ \cline{1-3} \cline{5-7} \cline{9-11}
     &          &        & & 3    & 0.132115    & 41221  & & 6   & 0.1282  & 98042   \\ \cline{1-3} \cline{5-7} \cline{9-11}
     &          &        & & 4    & 0.13014328  & 47419  & & 8   & 0.12739 & 99255   \\ \cline{1-3} \cline{5-7} \cline{9-11}
     &          &        & & 5    & 0.1290368   & 41440  & &     &         &         \\ \cline{1-3} \cline{5-7} \cline{9-11}
     &          &        & & 8    & 0.1274578   & 7116   & &     &         &         \\ \cline{1-3} \cline{5-7} \cline{9-11}
\end{tabular}
\caption{Parameter $\kappa$ used in the simulations at each $\beta_L=4/g_0^2$ and the number of measurements performed on the largest lattice.}
\label{table:kappa}
\end{table}

\section{Improvement in perturbative analysis}

The degree of improvement obtained with the clover term and boundary counter terms can be quantified with one loop order
perturbative analysis of the step scaling function \nr{stepscaling}.
Because the gauge action is identical with improved and unimproved fermions, to this order it is sufficient to consider only the fermion 
contribution to the step scaling.  To one loop order the step scaling function \nr{stepscaling} can be expanded as
\begin{eqnarray}
\Sigma(u,s,L/a) &=& g^2(g_0,sL/a)\vert_{g^2(g_0,L/a)=u} \nonumber \\
&=& u+(\Sigma_{1,0}+\Sigma_{1,1} N_f)u^2.
\end{eqnarray}
The fermion contribution is denoted by $\Sigma_{1,1}$. To evaluate these 
perturbative contributions we use the methods in 
\cite{Sint:1995ch,Sommer:1997jg}, and choose $s=2$.  The continuum limit of $\Sigma_{1,1}$ is given by the fermionic contribution to the one loop coefficient $b_{0}=\beta_0/(16\pi^2)$ of the beta function, i.e. 
\begin{equation}
\lim_{L/a\rightarrow 0}\Sigma_{1,1}=2 b_{0,1}\ln{2},
\end{equation}
where $b_{0,1}=1/(24\pi^2)$.

\begin{figure}
\centering
\includegraphics[scale=0.35]{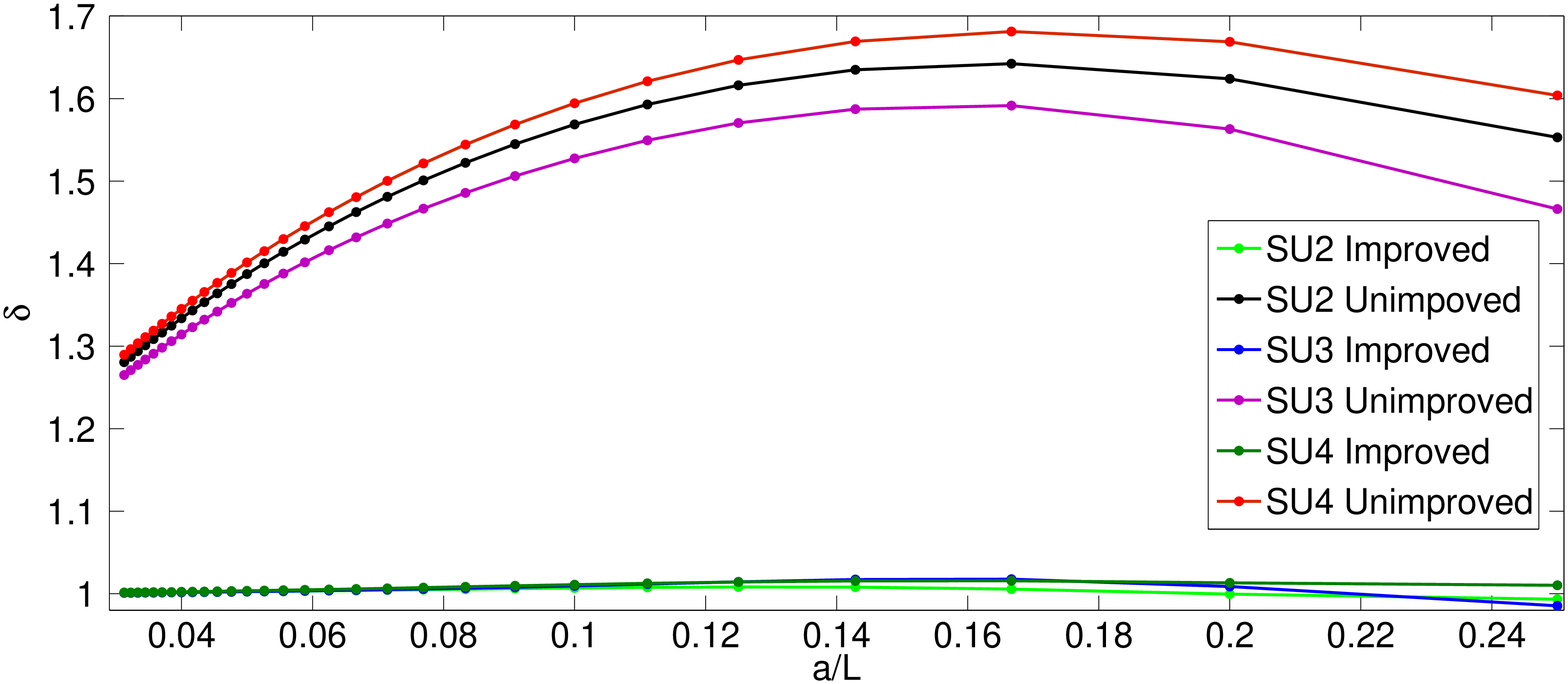}
\caption{Contribution of a massless Wilson quark to the step scaling function normalized to its continuum value at one loop order in perturbation theory. The top three curves show the result for gauge groups SU(4), SU(3) and SU(2) (from top to bottom) for unimproved Wilson fermions, while the lower three curves show the result after ${\cal{O}}(a)$ improvement has been taken into account.} 
\label{pertstep}
\end{figure}

In figure \ref{pertstep} we show $\delta\equiv \Sigma_{1,1}/(2 b_{0,1}\ln{2})$ both for unimproved Wilson fermions and with
${\cal{O}}(a)$ improvement. One immediately observes that without improvement,
$\Sigma_{1,1}$ depends strongly on $L/a$ and approaches the continuum limit 
only for presently impractically large lattices, while with improvement the large lattice artefacts are absent.  While this level of
improvement is probably not preserved at higher orders, this nevertheless strongly motivates the use improved actions in the lattice 
studies of these theories with Wilson fermions.  

\section{Simulations}
\label{measurements}

We use hybrid Monte Carlo (HMC) simulation algorithm with 2nd order
Omelyan integrator \cite{Omelyan,Takaishi:2005tz} and chronological initial values for the fermion matrix inversions \cite{Brower:1995vx}.  The trajectory length is 1, and the step length is tuned to have acceptance rate larger than 80\%.

\begin{table}
\centering
\begin{tabular}{|l|l|l|l|l|}
\hline
$\beta_L$ & $L/a=6$ & $L/a=8$ &  $L/a=12$ & $L/a=16$ \\
\hline
4   &1.2394(18)& 1.263(3) & 1.300(3) & 1.32(1) \\
3   & 1.832(5) & 1.882(5) & 1.971(18)& 2.02(2) \\
2.4 & 2.629(7) & 2.767(15)& 2.94(2)  & 3.17(4) \\
2.2 & 3.113(7) & 3.29(2)  & 3.58(3)  & 3.88(9) \\
2   & 3.93(2)  & 4.24(3)  & 4.77(8)  & 5.18(11)\\
1.9 & 4.65(2)  & 4.95(5)  & 5.48(7)  & 6.9(3)  \\
1.8 & 5.78(4)  & 6.43(7)  & 8.15(17) & 9.0(5)  \\
\hline
\end{tabular}
\caption{The measured values of $g^2$ at each $\beta_L=4/g_0^2$ and $L/a$ with $4$
 flavours of fermions.}
\label{table:couplingnf4}
\end{table}

\begin{table}
\centering
\begin{tabular}{|l|l|l|l|l|l|}
\hline
$\beta_L$ & $L/a=6$ & $L/a=8$ & $L/a=10$ &  $L/a=12$ & $L/a=16$ \\
\hline
8    & 0.5207(8)  & 0.5222(9)&         & 0.5274(13)& 0.528(4) \\
5    & 0.8585(15) & 0.868(3) &         & 0.875(3)  & 0.889(4) \\
4    & 1.095(3)   & 1.109(2) & 1.112(4)& 1.122(8)  & 1.135(7) \\
3    & 1.535(8)   & 1.555(10)&         & 1.587(10) & 1.623(15) \\
2.4  & 2.030(8)   & 2.087(16)&         & 2.19(3)   & 2.25(4) \\
2    & 2.655(15)  & 2.84(6)  & 2.76(3) & 2.95(5)   & 3.1(2) \\
1.8  & 3.25(3)    & 3.33(4)  & 3.45(5) & 3.47(4)   & 3.57(11) \\
1.5  & 5.40(6)    & 5.59(6)  & 5.57(11)& 5.75(11)  & 6.12(13) \\
1.44 & 7.21(10)   & 7.11(15) & 7.2(3)  & 7.3(3)    & 7.5(2) \\
1.4  & 9.74(13)   & 9.82(13) & 10.2(3) & 9.8(3)    & 10.4(4) \\
1.39 & 11.48(16)  & 13.4(3)  &         & 13.5(6)   & 13.5(8) \\
\hline
\end{tabular}
\caption{The measured values of $g^2$ at each $\beta_L$ and $L/a$ with $6$
 flavours of fermions.}
\label{table:couplingnf6}
\end{table}

\begin{table}
\centering
\begin{tabular}{|l|l|l|l|l|}
\hline
$\beta_L$ & $L/a=6$ & $L/a=8$ &  $L/a=12$ & $L/a=16$ \\
\hline
8   & 0.4700(2)& 0.4706(4) & 0.4705(5) & 0.4707(10) \\
6   & 0.6148(3)& 0.6159(5) & 0.6180(9) & 0.6181(19) \\
4   & 0.8897(9)& 0.8897(13)& 0.895(4)  & 0.895(3) \\
3   &1.1528(16)& 1.156(3)  & 1.150(2)  & 1.146(4) \\
2   & 1.651(4) & 1.653(5)  & 1.637(6)  & 1.624(13)  \\
1.7 & 1.924(4) & 1.907(5)  & 1.905(11) & 1.896(13)  \\
1.5 & 2.183(3) & 2.137(7)  & 2.116(11) & 2.10(2)  \\
1.3 & 2.542(8) & 2.473(9)  & 2.382(11) & 2.37(2)  \\
1   & 4.03(2)  & 3.55(2)   & 3.23(3)   & 3.09(4)  \\
\hline
\end{tabular}
\caption[a]{The measured values of $g^2$ at each $\beta_L$ and $L/a$ with $10$
 flavours of fermions.}
\label{table:couplingnf10}
\end{table}

\begin{figure}
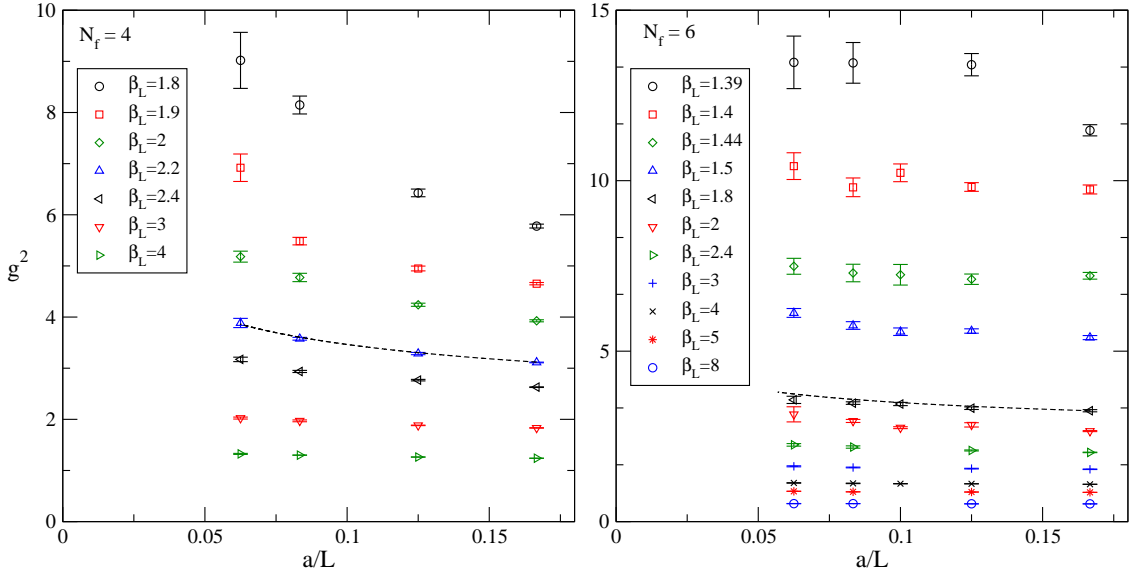

\centering
\includegraphics[height=0.5\textwidth]{nf4g2.eps}
\includegraphics[height=0.5\textwidth]{nf6g2.eps}
\caption[a]{
  The measured values of $g^2(g_0^2,L/a)$ against
  $a/L$ with 4 and 6 flavours of fermions.
  The black dashed line gives an example of the running in 2-loop perturbation theory at modest coupling, normalized so that it matches the measurement
  at $L/a=6$.
}
\label{fig:couplingnf46}
\end{figure}

\begin{figure}
\centering
\includegraphics[height=0.5\textwidth]{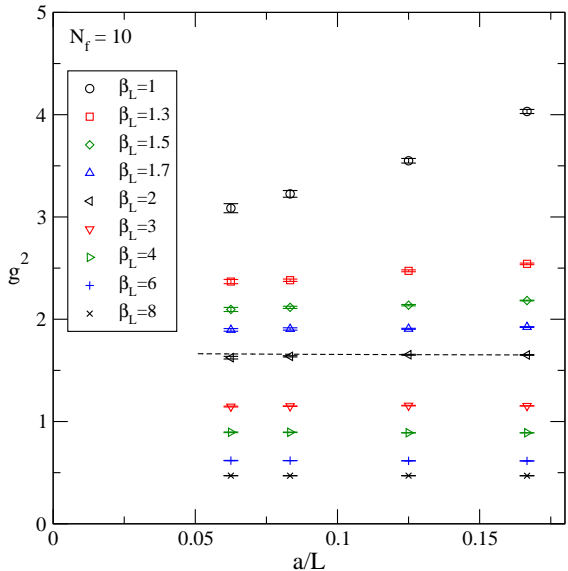}
\caption[a]{
  The measured values of $g^2(g_0^2,L/a)$ against
  $a/L$ with 10 flavours of fermions.
  The black dashed line gives an example of the running in 2-loop perturbation theory.
}
\label{fig:couplingnf10}
\end{figure}

The measured values of the running coupling squared in theories with 4, 6 and 10 fundamental representation fermions are given, respectively, in tables \ref{table:couplingnf4}, \ref{table:couplingnf6} and 
\ref{table:couplingnf10} and shown in figures \ref{fig:couplingnf46} and 
\ref{fig:couplingnf10}. The measurements are taken after every trajectory,
and the number of trajectories for each point is up to 230,000
(for $N_f=6$, volume $16^4$ and largest $\beta_L$-values).
For 6 and 10 flavours the strongest lattice couplings (smallest $\beta_L$) are very close to the strongest practical values for our action; at smaller $\beta_L$ the simulations
become either too slow or unstable, possibly signalling a proximity of a bulk phase transition.  These transitions are a lattice artefact and limit the range of allowed lattice couplings.

One can immediately recognize the main features from figures 
\ref{fig:couplingnf46} and \ref{fig:couplingnf10}:
at $N_f=4$, the coupling becomes stronger as lattice size increases, and the running becomes faster at stronger coupling.  This agrees with the expected QCD-like behaviour.  
In contrast, with 10 flavours we observe basically no running at all within our statistical errors, except with the strongest lattice coupling ($\beta_L=1$) used, where $g^2$ becomes smaller as volume increases.  This is the expected behaviour if we are at the strong coupling side of an infrared fixed point, where the $\beta$-function is positive.  However, at $\beta_L=1$ the data differ qualitatively from the larger $\beta_L$ measurements, possibly a signal of contamination from finite lattice spacing effects at large bare coupling.

\begin{figure}
\centering
\includegraphics[height=0.5\textwidth]{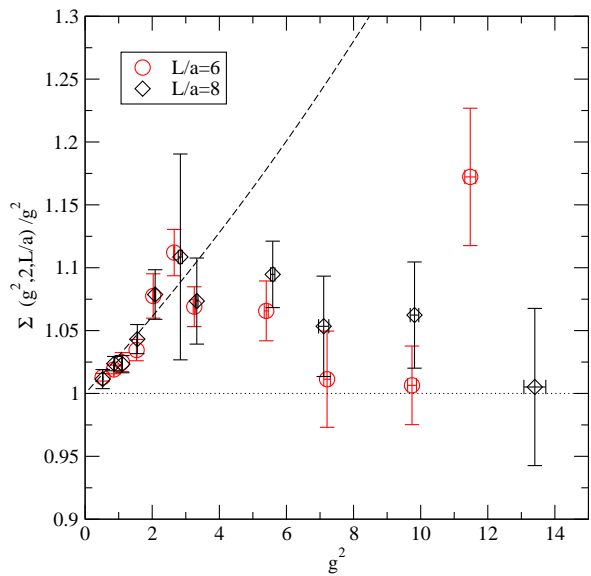}
\caption[stepscale]{The $N_f=6$ scaled lattice step scaling function
  $\Sigma(g^2,2,L/a)/g^2 = g^2(g_0^2,2L/a)/g^2(g_0^2,L/a)$ for
  $N_f=6$ calculated directly from the  data in table \ref{table:couplingnf6}.  The
  black dashed line is the continuum 2-loop perturbative result for
  $\sigma(g^2,2)$. }
\label{fig:step_nf6}
\end{figure}

In the theory with 6 fermions the running
remains small and slightly positive in the studied range of couplings. 
To illustrate the running we show the scaled step scaling function $\Sigma(g^2,2,L/a)/g^2 = g^2(g_0^2,2L/a)/g^2(g_0^2,L/a)$ 
in figure \ref{fig:step_nf6} at $L/a=6$ and $8$.
The running is compatible with the perturbation theory at small $g^2$, but deviates from it at large coupling.  The scaled step scaling measurements decrease at $g^2 \gsim 4$ and may reach unity at $g^2 \gsim 10$, indicating an ultraviolet fixed point.  However, there appears to be a systematic difference between the $L/a=6$ and $8$ points at large $g^2$.  Thus, proper continuum limit is necessary.

We also note that at $L/a=6$ the very largest coupling $\beta_L=4/g_0^2 = 1.39$ point deviates substantially from other points. This is caused by the
finite size effects at the smallest volume, clearly visible in figure \ref{fig:couplingnf46}.

\begin{figure}
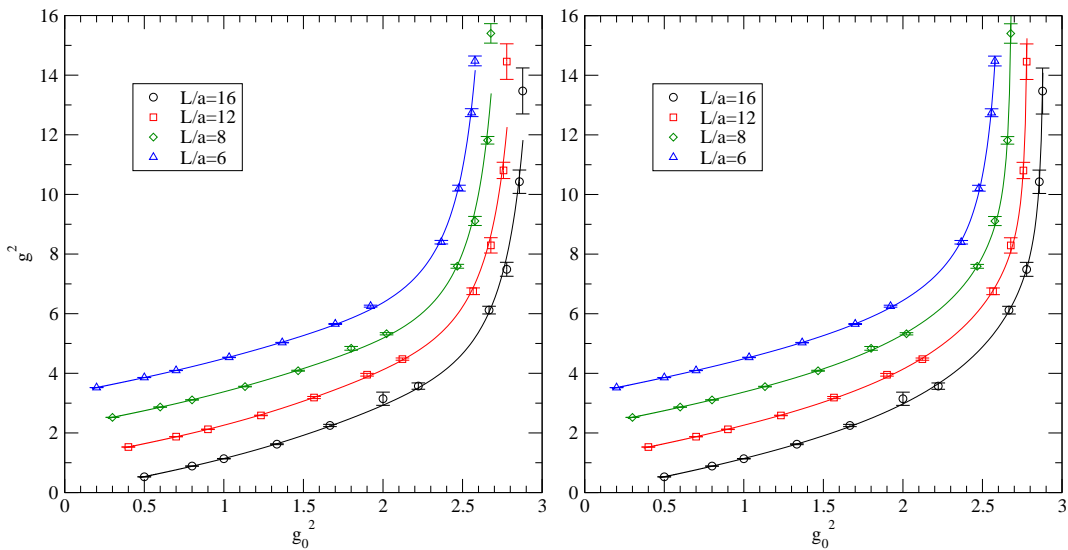

\centering
\includegraphics[scale=0.42]{betainterpolate}
\includegraphics[scale=0.42]{betaint2}
\caption[Interpolations in beta]{The interpolating functions for six fermions. The plot on the left shows the polynomial interpolation and the plot of the right shows the rational interpolation.
 For clarity, the graphs are displaced up by $3$,$2$,$1$ and $0$ and left by $0.3$,$0.2$,$0.1$ and $0$ for
 lattice sizes $L/a = 6$,$8$,$12$ and $16$ respectively.
}
\label{fig:betainterpolate}
\end{figure}


\begin{table}
\centering
\begin{tabular}{|l|l|l|l|l|l|l|l|l|}
\hline
$N_f$ & function 
  & $L/a=6$ & $L/a=8$ & $L/a=12$ & $L/a=16$ & d.o.f & $L/a=10$ & d.o.f \\ \hline
4 &
 Polynomial  & 1.565 & 6.063 & 18.961  & 3.854 & 3 & & \\
  &
 Rational    & 2.512 & 3.623 & 5.175   & 3.771 & 3 & & \\ \hline
6 &
 Polynomial  & 35.25  & 72.76 & 30.20 & 16.11  & 6 & & \\
  &
 Rational    & 8.030  & 5.897 & 6.968 & 4.661  & 7 & 3.545 & 2\\\hline
10 & Polynomial  & 13.744 & 8.087 & 13.802 & 5.732 & 5 & & \\
   &
   Rational    & 9.896  & 6.271 & 12.704 & 6.301 & 6 & & \\
\hline
\end{tabular}
\caption[$\chi^2$ values]{$\chi^2$ values and free degrees of freedom in the polynomial fits \ref{eq:g2function} and the rational fits \ref{eq:g2function2}.}
\label{table:chi2}
\end{table}

The continuum limit of the step scaling function $\Sigma(g^2,2,L/a)$ has to be
evaluated at constant Schrödinger functional coupling $g^2$.  However, the
measurements were performed at selected fixed values of $g_0^2=4/\beta_L$,
which do not correspond to the same $g^2$-values at $L/a=6$ and $8$.  Therefore,
it is necessary to shift the measurements so that $g^2$-values match.
This could be done by performing new simulations so that $g^2$ at $L/a=6$
and $8$ match, or by reweighting in $g_0^2$ and $\kappa$.  However, a much more economical and convenient way to achieve this is to 
interpolate the actual measurements of $g^2(g_0^2,L/a)$ at each lattice size $L/a$ by fitting to a function of $g_0^2$.  This results in a
``measurement'' of $g^2(g_0^2,L/a)$ over a continuous range of $g_0^2$.

It is necessary that the measurements cover the interpolated range 
densely enough; otherwise the form of the fitted function ansatz may influence 
the final results.  Obviously, the fit must also be statistically good in order to describe the true behaviour of the data.
The interpolating function also somewhat averages out fluctuations in individual data points.  

We have done the interpolation using polynomial and rational interpolating functions.  Polynomial function is a power series in the bare coupling, and it is commonly used in the literature \cite{Appelquist:2009ty,Bursa:2010xr}:
\begin{align}
  \frac{1}{g^2(g_0^2,L/a)} = \frac{1}{g_0^2} \left[ 1 + \sum_{i=1}^n
  c_i g_0^{2i}\right]. \label{eq:g2function}
\end{align}
In the case of 4 and 10 flavours we truncate the series
at $n=4$ and in the case of 6 flavours, where more $\beta_L$-values are available, at $n=5$.
These values were chosen to 
optimise the confidence levels of the fits, while keeping
the number of the fit parameters tolerably small.
We also kept the
same number of terms for each $L/a$.  The $\chi^2$-values of the fits
are shown in table \ref{table:chi2}, and, for $N_f=6$, the fit is shown
in figure \ref{fig:betainterpolate}, left.

It is clear that the polynomial fits do not fit the data very well,
especially at the strongest bare couplings where the measured
$g^2$ increases very rapidly.  This leads us to try a rational function interpolation:
\begin{align}
  \frac{1}{g^2(g_0^2,L/a)} = \frac{1}{g_0^2} \left[\frac{ 1 + \sum_{i=1}^n
  a_i g_0^{2i}}{1 + \sum_{i=1}^m
  b_i g_0^{2i}}\right]. \label{eq:g2function2}
\end{align}
For 4 and 6 fermions the number of terms were chosen to be $n=m=2$ and for 10 fermions $n=1$ and $m=2$.\footnote{%
For $N_f=10$ choosing $n=2$ produces a singularity within the fitting range,
which is not acceptable.}
Again, these parameters were chosen to optimize the confidence level
without producing singularities within the range of $g_0^2$ of the measurements. 
The stability of the fits was checked by varying $n$ or $m$.  
The rational interpolation captures the rapid increase at strong bare couplings better, and the confidence levels of the fits are better in all cases, as shown in table \ref{table:chi2} and figure \ref{fig:betainterpolate}.  Thus, we perform the subsequent analysis using the rational interpolation functions.



\begin{figure}
\centering
\includegraphics[scale=0.45]{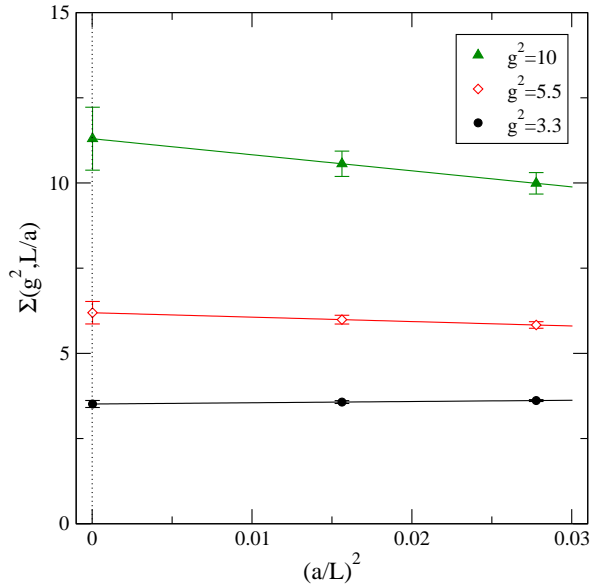}
\caption[Continuum extrapolation of sigma.]{ The step scaling function extrapolated to the continuum limit with $N_f=6$ and for three chosen values of $u=g^2$. }
\label{sigmacontNf6}
\end{figure}

The interpolating functions are used to calculate $\Sigma(u,s,L/a)$ with $L/a=6,8$ in a continuous range of values of $u=g^2$. This enables us to perform a continuum  extrapolation.  Since we expect most of the order $a$ errors to have been removed, we fit the data
at $L/a=6$ and $L/a=8$ with a function of the form
\begin{align}
\Sigma(u,2,L/a) =  \sigma(u,2) + c(u) \left ( L/a \right )^{-2}
\label{continuum}
\end{align}
independently at each value of $u$.
This extrapolation is shown for $N_f=6$ for three selected values of $u$ in figure \ref{sigmacontNf6}.
To estimate the systematic errors from the extrapolation we compare with the result
obtained by simply taking the largest volume step scaling, 
$\sigma(u,2) = \Sigma(u,2,8)$.

Because we only have step scaling data at $L/a=6$ and $8$, we are naturally not able to verify that the first order in $L/a$ -term is indeed small or that the higher order contributions can be neglected.  However, we note that using a first order only extrapolation would give results roughly comparable with the extrapolation (\ref{continuum}), but with larger statistical errors and further away from the largest volume result.
The $L/a=10$ results with $N_f=6$ are not used for the measurement of the step scaling for the running coupling, but are needed for the measurement of the anomalous dimension.

\begin{figure}
\centering
\includegraphics[scale=0.5]{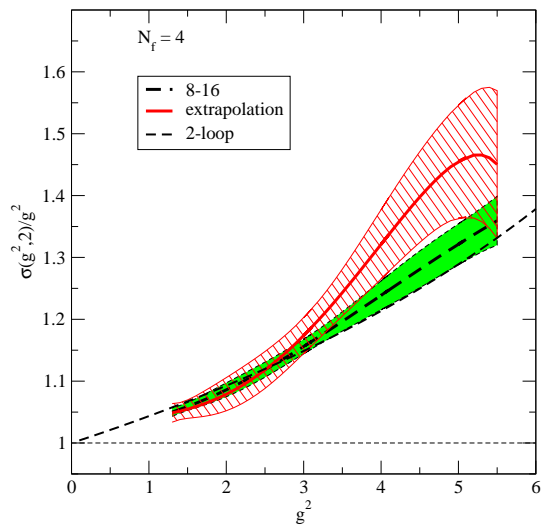}
\caption[The step scaling function with 4 fermions.]{
  (colour online) The scaled step scaling function $\sigma(g^2,2)/g^2$ with 4 fermions.
  The thick red line corresponds to the continuum extrapolation,
  \eq{continuum}, and the hashed band to the statistical errors of the
  extrapolation.  The thick dashed line with the shaded error band is the
  largest volume step scaling function without extrapolation.  The thin
  dashed line is the 2-loop perturbative value of $\sigma(g^2,2)/g^2$.
}
\label{sigmanf4}
\end{figure}

\begin{figure}
\centering
\includegraphics[scale=0.5]{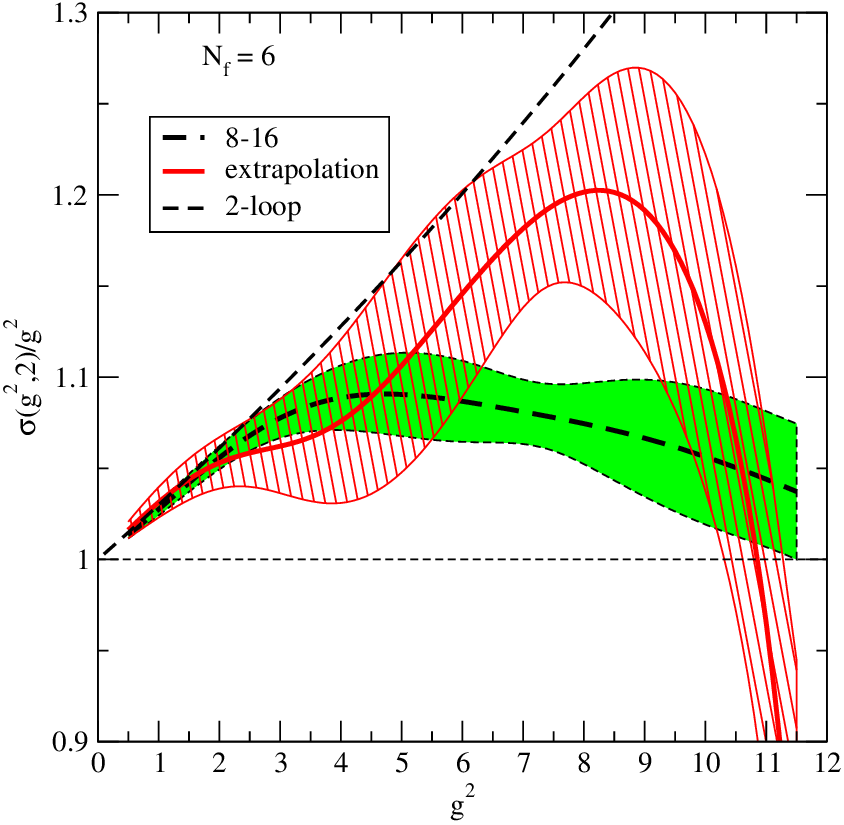}
\caption[The step scaling function with 6 fermions.]{
  As in figure \ref{sigmanf4} but with 6 fermion flavours.}
\label{sigmanf6}
\end{figure}

\begin{figure}
\centering
\includegraphics[scale=0.5]{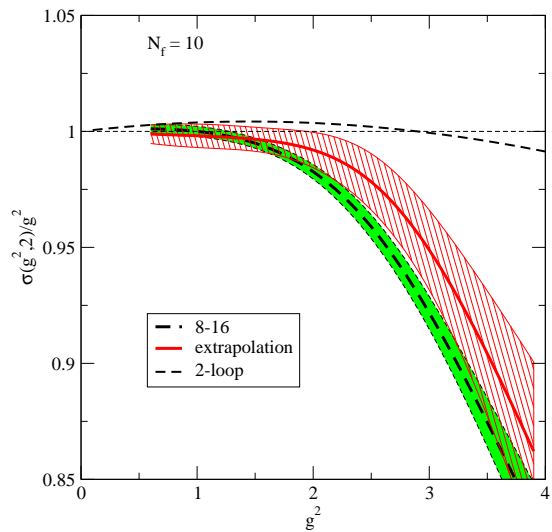}
\caption[The step scaling function with 10 fermions.]{
  As in figure \ref{sigmanf4} but with 10 fermion flavours.}
\label{sigmanf10}
\end{figure}

The final results of the step scaling functions are shown 
in figures \ref{sigmanf4}, \ref{sigmanf6} and \ref{sigmanf10} for 4, 6 and 10 fermion flavours respectively. The errors shown include only the statistical errors from the measurements, fits and extrapolation. A rough measure of the systematic errors can be obtained from the comparison of the infinite volume extrapolated result and the result using only the largest volume step scaling without extrapolation. The error propagation is calculated using jackknife blocking throughout the whole analysis.

\begin{figure}
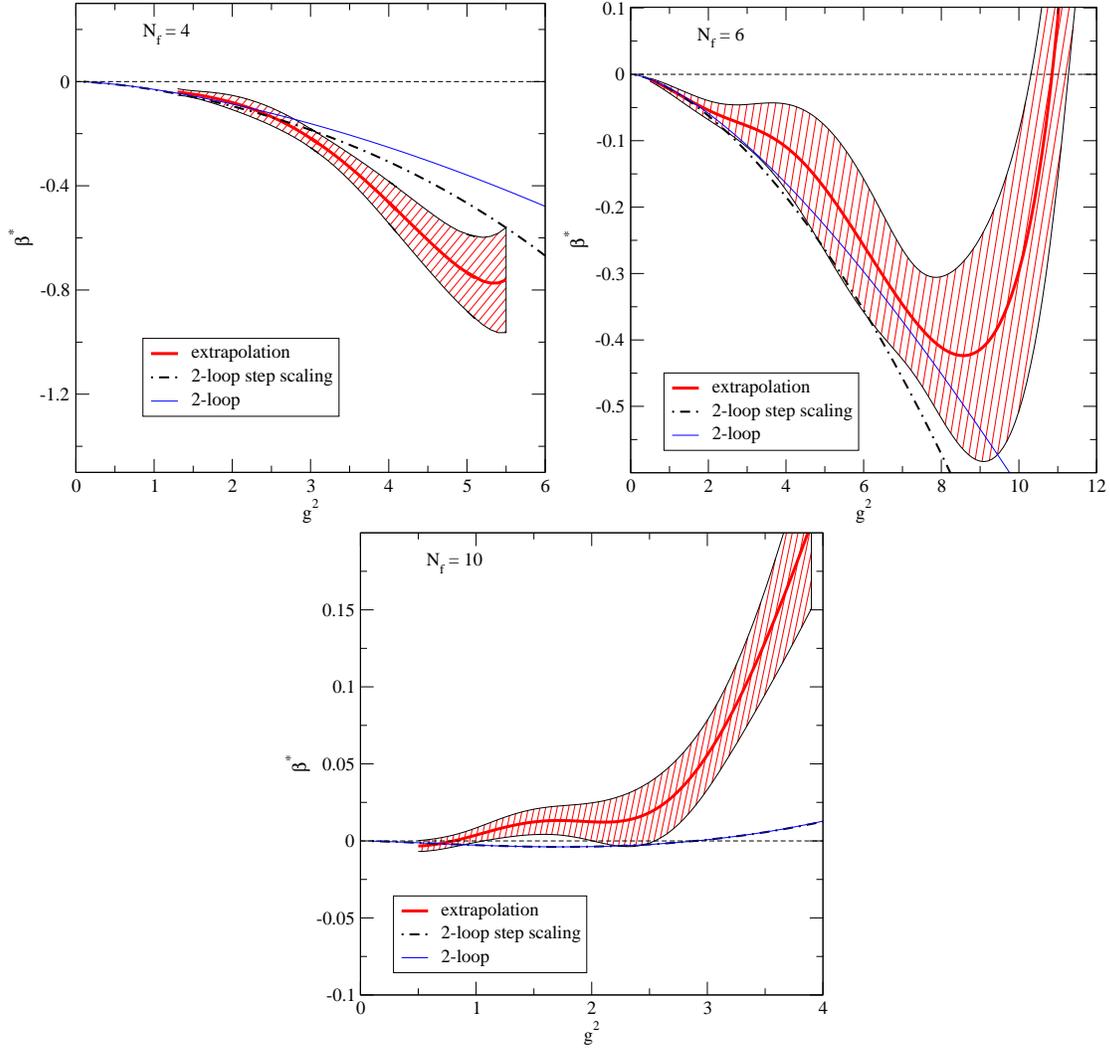

\centering
\includegraphics[width=0.48\textwidth]{nf4bf.eps}
\includegraphics[width=0.48\textwidth]{nf6bf.eps}\\
\includegraphics[width=0.48\textwidth]{nf10bf.eps}
\caption[The estimators for the $\beta$-function.]{ The estimators for the $\beta$-function in equation \eq{eq:beta*}.
}
\label{fig:beta*}
\end{figure}

We have also calculated the estimator for the $\beta$-function, $\beta^*$, defined in \eq{eq:beta*}.
These are shown in figure \ref{fig:beta*}, together with the 2-loop $\beta$-function and  $\beta^*$ obtained from the 2-loop perturbative step scaling function. Thus, the difference between the perturbative curves gives a measure of the error made in the approximation in \eq{eq:beta*}.

What can we conclude from the results?
From figure \ref{sigmanf4} we can observe that, in the range of couplings studied, the $N_f=4$ case behaves as expected: $\beta$-function is negative and becomes smaller as the coupling increases, and agrees overall with the 2-loop perturbative $\beta$-function. Naturally deviations from the perturbation theory are expected at larger couplings. However, for our purposes it is sufficient to verify the QCD-like behaviour and we did not attempt to reach smaller couplings.

For $N_f=6$, the results 
in figure \ref{sigmanf6} show that the largest volume step scaling deviates from the perturbative result already at $g^2 \sim 4$, after which it decreases.  However, the continuum extrapolation remains close to the perturbative curve up to $g^2\sim 7$--$8$.
After this the curve turns sharply downwards; this behaviour is caused by the anomalous $L/a=6$, $\beta_L=1.39$ point visible in figures \ref{fig:couplingnf46} and
\ref{fig:step_nf6}.  The largest volume step scaling (shaded band) can be compared to the corresponding $L/a=8$ uninterpolated step scaling in figure \ref{fig:step_nf6}.  

The large difference between the largest volume and extrapolated step scaling functions is naturally due to the fact that the step scaling $\Sigma(g^2,L/a)$ at $L/a=6$ and $L/a=8$ differ substantially, and the lever arm to the continuum ($a/L\rightarrow 0$) is long.  The variation between these two
curves gives an estimate of the systematic errors in the extrapolation.
The largest values of the Schr\"odinger functional coupling we achieved was around $g^2 \approx 12$ ($\alpha \approx 0.96$).  Unfortunately, neither the coupling is strong enough nor the errors are sufficiently small in order to unambiguously distinguish between an infrared fixed point at $g^2\gsim 11$ and ``walking'' behaviour, where the step scaling function starts to increase again at stronger coupling.
With our choice of the action we cannot reach stronger couplings because simulations become rapidly impractical at smaller $\beta_L$ than 1.39, our
strongest lattice coupling at $N_f=6$.  Obviously, calculating the step scaling at volumes $L/a=10$ and $20$ (or larger) would be needed in order to stabilize the continuum extrapolation.

Finally, for $N_f=10$ the measured $\beta$-function is essentially compatible with zero or with the perturbative $\beta$-function at $g^2 < 2.5$.  In this case the 2-loop perturbative $\beta$-function is expected to be relatively accurate (higher order perturbative corrections in MS-scheme are small \cite{vanRitbergen:1997va}).  The value of the $\beta$-function is very small, and clearly, the accuracy of our results falls far short from being able to resolve the behaviour of the $\beta$-function in this range.  Nevertheless, theoretically we know that the $\beta$-function must be negative at small coupling.  At $g^2 \gsim 2.5$ the measured $\beta$-function deviates significantly upwards from the perturbative one.  Combined with the knowledge that the $\beta$-function is negative at small couplings, this indicates the presence of an infrared fixed point.  The onset for this behaviour is very close to the fixed point in the
2-loop $\beta$-function.
This deviation is in practice caused by the strongest coupling $\beta_L=1$ data from the lattice: from figure \ref{fig:couplingnf10} we can see that only at $\beta_L=1$ the measured coupling $g^2 > 2.5$.  At these couplings we can expect strong lattice artefacts; thus, we believe that the deviation from the perturbative value is caused by unaccounted for systematic errors in the continuum extrapolation.
We still emphasize that in absolute numbers the deviation is still rather small;
in order to resolve very slow running we would need extremely accurate measurements with correspondingly very small systematic errors.  This does not appear to be doable at $N_f=10$ with the methods used here.

\begin{table}
\centering
\begin{tabular}{|l|l|l|l|l|l|l|}
\hline
$\beta_L$ & $L/a=6$ & $L/a=8$ & $L/a=10$ &  $L/a=12$ & $L/a=16$ & $L/a=20$ \\
\hline
2.4  & 0.9666(8) & 0.9353(9) & 0.9171(19)& 0.9014(19)& 0.8870(15)& 0.865(4) \\  
2    & 0.8953(10)& 0.857(3)  & 0.838(2)  & 0.823(3)  & 0.793(4)  & 0.766(6) \\
1.5  & 0.702(4)  & 0.669(3)  & 0.646(4)  & 0.618(6)  & 0.586(5)  & 0.573(6) \\
1.44 & 0.636(3)  & 0.610(3)  & 0.588(4)  & 0.572(4)  & 0.548(4)  & 0.517(6) \\
1.4  & 0.543(5)  & 0.547(5)  & 0.539(5)  & 0.534(5)  & 0.508(6)  & 0.480(8) \\
1.39 & 0.508(5)  & 0.515(7)  & 0.520(6)  & 0.517(6)  & 0.488(8)  & 0.476(9) \\
\hline
\end{tabular}
\caption{The measured values of $Z_P$ at each $\beta_L$ and $L$ with $6$
 flavours of fermions.}
\label{table:zpnf6}
\end{table}

We measure the anomalous dimension for the interesting case of SU(2) with 6 flavours of fermions using the mass step scaling method described in section \ref{model}.  This method is much less noisy than the measurement of the coupling, and we can now reliably use lattices of size $20^4$.  The measurements of $Z_p$ are listed in table \ref{table:zpnf6}.

\begin{table}
\centering
\begin{tabular}{|l|l|l|l|l|l|l|}
\hline
   $L/a=6$ & $L/a=8$ & $L/a=10$ & $L/a=12$ & $L/a=16$ & $L/a=20$  \\ \hline
25.15 & 6.48 & 0.28 & 0.84 & 3.00 & 1.50 \\
\hline
\end{tabular}
\caption[$\chi^2$ values]{$\chi^2$ values (2 degrees of freedom) for the $Z_P$ fits
  for $N_f=6$ using \eq{eq:zpfunction}.  }
\label{table:zpchi2}
\end{table}

\begin{figure}
\centering
\includegraphics[scale=0.42]{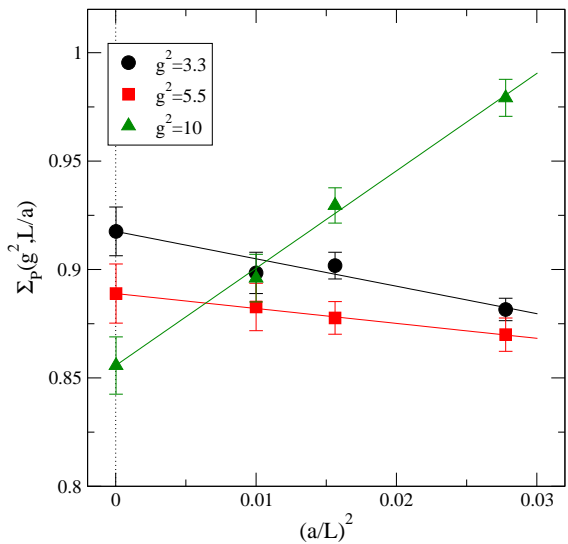}
\caption[a]{The mass step scaling function extrapolated to the continuum limit, using $N_f=6$ data and shown for three chosen values of $g^2$. }
\label{sigmapcontNf6}
\end{figure}

\begin{figure}
\centering
\includegraphics[width=0.5\textwidth]{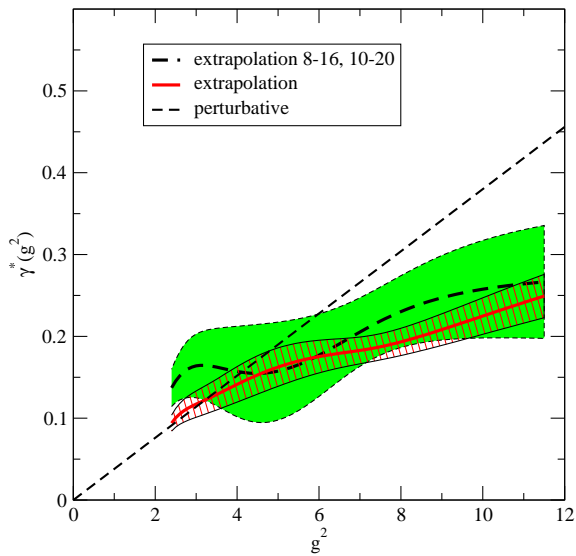}
\caption[The mass anomalous dimension with 6 fermions.]{The estimate of the mass anomalous dimension, $\gamma^*(g^2)$, with $N_f=6$.  Shown are the continuum extrapolation using the mass step scaling $\Sigma(L/a,2,g^2)$ with all three volumes $L/a=6,8,10$ (hashed band) and only with $L/a = 8,10$ (shaded band).}
\label{figgamma}
\end{figure}

We start the analysis by finding an interpolating function to the measured values of $Z_P$ by fitting with a power series
\begin{align}
  Z_P(\beta_L,L/a) = 1 + \sum_{i=1}^n
  c_i  g_0^{2i}, \label{eq:zpfunction}
\end{align}
where we have truncated the series at $n=4$. The $\chi^2$ values for the fits are given in table \ref{table:zpchi2}.
The fit at $L/a=6$ has bad $\chi^2/$ value; however, its effect to the final extrapolation is small. 
We check the systematic errors of the interpolation by also truncating at $n=3$; the results remain essentially the same with somewhat increased statistical errors.
From the interpolated $Z_p(\beta_L,L/a)$ we obtain the mass step scaling function $\Sigma_P(u,s,L/a)$ at $L/a=6,8$ and $10$ using \eq{Sigmap}, using $u=g^2$ from rational fit in \eq{eq:g2function2}. The continuum extrapolation is then done by fitting to the extrapolating
function
\begin{align}
\Sigma_P(u,2,L/a) =  \sigma_P(u,2) + c(u) \left ( L/a \right )^{-2}.
\end{align}
The fit is shown in figure \ref{sigmapcontNf6}.
We check the systematic effects in the extrapolation by leaving out the smallest volume step scaling at $L/a=6$.  The results remain compatible with each other, but with naturally much smaller final statistical errors when $L/a=6$ is included.  Step scaling function is converted to the estimate of the anomalous dimension using \eq{gammastar}, and the results are shown in figure \ref{figgamma}.  

In this case the non-trivial continuum extrapolation is essential in order to 
find the monotonic growth of $\gamma^*(g^2)$.  If we would take e.g. $L/a=8$ step scaling alone to estimate the continuum behaviour, $\gamma^*$ would decrease at $g^2 \gsim 6$.  This emphasizes the importance of the controlled continuum limit.

The mass anomalous dimension we measure is somewhat smaller than the perturbative one at strong coupling.  It remains small 
at all measured values of the coupling, and if there is a fixed point at $g^2 \gsim 12$, 
the anomalous dimension is $\gsim 0.25$.  

The behaviour of the anomalous dimension is in contrast with the results obtained with an unimproved Wilson action \cite{Bursa:2010xr}, where $\gamma^*$ significantly larger than the perturbative one was seen already at $g^2 \approx 5$, albeit with large errors.  Both of the
measurements were done using the Schr\"odinger functional scheme, and thus should yield identical results in the continuum limit.  
This again points to the importance of controlling the cutoff effects and the continuum extrapolation.  On the other hand, the 
behaviour we measure is qualitatively similar to the one observed in SU(2) gauge theory with adjoint fermions using HYP-smeared
improved fermions \cite{DeGrand:2011qd}.  

For the $N_f=10$ case we did not perform a measurement of the mass anomalous dimension. Since our results for the coupling constant in the $N_f=10$ case are compatible with perturbative results, we expect this to hold also for the mass anomalous dimension. In other words, we expect that $\gamma^\ast=0.08$ corresponding with a perturbative fixed point at $g^2\simeq 2.90$.

\section{Conclusions}
\label{conclusions}

In this paper we have presented the results of a lattice study of SU(2) gauge theory with $N_f = 4$, $6$ or $10$ flavours of Wilson fermions in the
fundamental representation of the gauge group.  The numbers of fermion flavours were chosen according to the expected boundaries of the conformal window: $N_f=10$ is the largest value where the theory is still asymptotically free, and hence it has a fixed point at small coupling where perturbation theory is expected to be accurate.  On the other hand, $N_f=6$ is expected to be close to the lower boundary of the conformal window, whereas $N_f=4$ should be safely below it and hence QCD-like.

We have measured the $\beta$-functions of the above theories using the Schr\"odinger functional scheme.  We also measure the mass anomalous dimension $\gamma$ for the $N_f=6$ theory.
In our analysis we have used perturbatively improved Wilson-clover action, together with the improvement of the boundary terms in the Schr\"odinger functional approach.  With this improvement we expect most of the $O(a)$ errors to be eliminated.

Overall our findings agree with the expectations.  For $N_f=10$ we observe a very small $\beta$-function at small couplings, and a positive $\beta$-function at larger ($g^2\gsim 3$) couplings.  Together with the fact that due to the asymptotic freedom the $\beta$-function must be negative at small enough coupling, this indicates that the $\beta$-function has a zero somewhere at $g^2 \lsim 3$.  This agrees with the perturbative  2-loop $\beta$-function which vanishes at $g^2 \approx 2.90$.

At the interesting case of six flavours we observe step scaling (or $\beta$-function) behaviour compatible with the perturbation theory up to $g^2 \approx 5$.  Above this the $\beta$-function starts to approach zero, with a possible fixed point around $g^2 \gsim 12$.  However, our statistical errors are too large in order to fully resolve the behaviour of the $\beta$-function: instead of the fixed point, the $\beta$-function could as well start to decrease again at stronger couplings.  This kind of behaviour is characteristic for the walking coupling.  It would be very interesting to resolve the behaviour in this case: as discussed in the introduction, in the ladder approximation the chiral symmetry breaking sets in around the critical coupling $g^2_c \sim 4\pi^2/(3C_2(R)) \approx 17$.  The estimated provisional fixed point is close to this value (although computed with a different schema), and thus it is indeed plausible that either the fixed point or the chiral symmetry breaking is realized.

Obviously, simulation at much larger volumes (closer to the continuum) would be required to fully pin down the behaviour of the $\beta$-function.  However, it is just barely possible to reach these values of the Schr\"odinger functional scheme coupling with the action we use: in order to reach strong
Schr\"odinger functional coupling one has to use strong bare lattice coupling, and the couplings used in this work are already near the largest values which can be used in practice.  Clearly, it is important to develop an action which can be used at stronger couplings.

We also measured the mass anomalous dimension using six fermion flavours.  The dimension is seen to grow slowly as the coupling increases; somewhat slower than the perturbative result at strong couplings, but overall the result in this case is as expected.

Finally, for four flavours the observed behaviour is unsurprising: the measurements indicate a smoothly decreasing $\beta$-function and QCD-like behaviour.

\acknowledgments 
This work is supported by the Academy of Finland
grant 114371.  JR acknowledges the support from Väisälä foundation and
TK from Magnus Ehrnrooth foundation.  The simulations were performed at the Finnish IT
Center for Science (CSC), Espoo, Finland, and at 
EPCC, University of Edinburgh. 
Parts of the simulation program have been derived from
the MILC lattice simulation program \cite{MILC}.


\begin{thebibliography}{199}

\bibitem{Georgi:2007ek}
  H.~Georgi, {\em{Unparticle Physics}},
  Phys.\ Rev.\ Lett.\  {\bf 98}, 221601 (2007)
  [arXiv:hep-ph/0703260].

\bibitem{Georgi:2007si}
  H.~Georgi, {\em{Another Odd Thing About Unparticle Physics}},
  Phys.\ Lett.\  B {\bf 650}, 275 (2007)
  [arXiv:0704.2457 [hep-ph]];
\bibitem{Cheung:2007zza}
  K.~Cheung, W.~Y.~Keung and T.~C.~Yuan,
  {\em{Collider signals of unparticle physics}},
  Phys.\ Rev.\ Lett.\  {\bf 99}, 051803 (2007)
  [arXiv:0704.2588 [hep-ph]].
  
\bibitem{Sannino:2008nv}
  F.~Sannino and R.~Zwicky,
  {\em Unparticle \& Higgs as Composites,}
  Phys.\ Rev.\  D {\bf 79} (2009) 015016
  [arXiv:0810.2686 [hep-ph]].

\bibitem{TC} 
S.~Weinberg,
{\em{Implications Of Dynamical Symmetry Breaking: An Addendum}},
Phys.\ Rev.\ D {\bf 19}, 1277 (1979);
L.~Susskind,
{\em{Dynamics Of Spontaneous Symmetry Breaking In The Weinberg-Salam Theory}},
Phys.\ Rev.\ D {\bf 20}, 2619 (1979).

\bibitem{Eichten:1979ah}
  E.~Eichten and K.~D.~Lane,
  {\em{Dynamical Breaking Of Weak Interaction Symmetries}},
  Phys.\ Lett.\  B {\bf 90}, 125 (1980).

\bibitem{Hill:2002ap}
  C.~T.~Hill and E.~H.~Simmons,
  {\em Strong dynamics and electroweak symmetry breaking,}
  Phys.\ Rept.\  {\bf 381}, 235 (2003)
  [Erratum-ibid.\  {\bf 390}, 553 (2004)]
  [arXiv:hep-ph/0203079].

\bibitem{Sannino:2008ha}
  F.~Sannino, 
  {\em{Dynamical Stabilization of the Fermi Scale: Phase Diagram of Strongly
      Coupled Theories for (Minimal) Walking Technicolor and Unparticles}},
 arXiv:0804.0182 [hep-ph].

\bibitem{Sannino:2004qp}
  F.~Sannino and K.~Tuominen,
  Phys.\ Rev.\  D {\bf 71}, 051901 (2005)
  [arXiv:hep-ph/0405209].

\bibitem{Banks:1981nn}
  T.~Banks and A.~Zaks,
  {\em On the Phase Structure of Vector-Like Gauge Theories with Massless
    Fermions,}
  Nucl.\ Phys.\  B {\bf 196}, 189 (1982).

\bibitem{Appelquist:1986an}
  T.~W.~Appelquist, D.~Karabali and L.~C.~R.~Wijewardhana,
  Phys.\ Rev.\ Lett.\  {\bf 57}, 957 (1986).

\bibitem{vanRitbergen:1997va}
  T.~van Ritbergen, J.~A.~M.~Vermaseren and S.~A.~Larin,
  Phys.\ Lett.\  B {\bf 400}, 379 (1997)
  [arXiv:hep-ph/9701390].




\bibitem{Bursa:2010xr}
  F.~Bursa, L.~Del Debbio, L.~Keegan, C.~Pica and T.~Pickup,
  {\em Mass anomalous dimension and running of the coupling in SU(2) with six
    fundamental fermions},
  arXiv:1010.0901 [hep-ph].

\bibitem{Ohki:2010sr}
  H.~Ohki {\it et al.},
  {\em Study of the scaling properties in SU(2) gauge theory with eight flavors,}
  arXiv:1011.0373 [hep-lat].

\bibitem{Voronov}
  G.~Voronov {\it et al.} (Lattice strong dynamics collaboration):
  {\em Lattice Study of the Conformal Window in Two-Color Yang-Mills Theory},
  Talk at LATTICE 2011, Squaw Valley, July 12 2011.



\bibitem{Catterall:2007yx}
  S.~Catterall and F.~Sannino,
  {\em{Minimal walking on the lattice}},
  Phys.\ Rev.\  D {\bf 76}, 034504 (2007)
  [arXiv:0705.1664 [hep-lat]].

\bibitem{Hietanen:2008mr}
  A.~J.~Hietanen, J.~Rantaharju, K.~Rummukainen and K.~Tuominen,
  {\em Spectrum of SU(2) lattice gauge theory with two adjoint Dirac flavors,}
  JHEP {\bf 0905}, 025 (2009)
  [arXiv:0812.1467 [hep-lat]]

\bibitem{DelDebbio:2008zf}
  L.~Del Debbio, A.~Patella and C.~Pica,
  {\em{Higher representations on the lattice: numerical simulations.
      SU(2) with adjoint fermions}},
  Phys.\ Rev.\  D {\bf 81} (2010) 094503
  [arXiv:0805.2058 [hep-lat]].

\bibitem{Catterall:2008qk}
  S.~Catterall, J.~Giedt, F.~Sannino and J.~Schneible,
  {\em Phase diagram of SU(2) with 2 flavors of dynamical adjoint quarks,}
  JHEP {\bf 0811}, 009 (2008)
  [arXiv:0807.0792 [hep-lat]].

\bibitem{Hietanen:2009az}
  A.~J.~Hietanen, K.~Rummukainen and K.~Tuominen,
  {\em Evolution of the coupling constant in SU(2) lattice gauge 
    theory with two adjoint fermions,}
  Phys.\ Rev.\  D {\bf 80}, 094504 (2009)
  [arXiv:0904.0864 [hep-lat]].

\bibitem{Bursa:2009we}
  F.~Bursa, L.~Del Debbio, L.~Keegan, C.~Pica and T.~Pickup,
  {\em Mass anomalous dimension in SU(2) with two adjoint fermions,}
  Phys.\ Rev.\  D {\bf 81}, 014505 (2010)
  [arXiv:0910.4535 [hep-ph]];

\bibitem{DelDebbio:2009fd}
  L.~Del Debbio, B.~Lucini, A.~Patella, C.~Pica and A.~Rago,
  {\em Conformal vs confining scenario in SU(2) with adjoint     fermions,}
  Phys.\ Rev.\  D {\bf 80}, 074507 (2009)
  [arXiv:0907.3896 [hep-lat]].

\bibitem{DelDebbio:2010hx}
  L.~Del Debbio, B.~Lucini, A.~Patella, C.~Pica and A.~Rago,
  {\em The infrared dynamics of Minimal Walking Technicolor,}
  Phys.\ Rev.\  D {\bf 82} (2010) 014510
  [arXiv:1004.3206 [hep-lat]].

\bibitem{DelDebbio:2010hu}
  L.~Del Debbio, B.~Lucini, A.~Patella, C.~Pica and A.~Rago,
  {\em Mesonic spectroscopy of Minimal Walking Technicolor,}
  Phys.\ Rev.\  D {\bf 82} (2010) 014509
  [arXiv:1004.3197 [hep-lat]].

\bibitem{Bursa:2011ru}
  F.~Bursa, L.~Del Debbio, D.~Henty, E.~Kerrane, B.~Lucini, A.~Patella, C.~Pica, T.~Pickup {\it et al.},
  {\em Improved Lattice Spectroscopy of Minimal Walking Technicolor,}
  [arXiv:1104.4301 [hep-lat]].

\bibitem{DeGrand:2011qd}
  T.~DeGrand, Y.~Shamir, B.~Svetitsky,
  {\em Infrared fixed point in SU(2) gauge theory with adjoint fermions,}  
  [arXiv:1102.2843 [hep-lat]].




\bibitem{Damgaard:1997ut}
  P.~H.~Damgaard, U.~M.~Heller, A.~Krasnitz and P.~Olesen,
  {\em On lattice QCD with many flavors,}
  Phys.\ Lett.\  B {\bf 400}, 169 (1997)
  [arXiv:hep-lat/9701008].

\bibitem{Appelquist:2007hu}
  T.~Appelquist, G.~T.~Fleming and E.~T.~Neil,
  {\em{Lattice Study of the Conformal Window in QCD-like Theories}},
  Phys.\ Rev.\ Lett.\  {\bf 100}, 171607 (2008)
  [arXiv:0712.0609 [hep-ph]].

\bibitem{Appelquist:2009ty}
  T.~Appelquist, G.~T.~Fleming and E.~T.~Neil,
  {\em Lattice Study of Conformal Behavior in SU(3) Yang-Mills Theories,}
  Phys.\ Rev.\  D {\bf 79}, 076010 (2009)
  [arXiv:0901.3766 [hep-ph]].

\bibitem{Fodor:2009wk}
  Z.~Fodor, K.~Holland, J.~Kuti, D.~Nogradi and C.~Schroeder,
  {\em Nearly conformal gauge theories in finite volume,}
  Phys.\ Lett.\  B {\bf 681}, 353 (2009)
  [arXiv:0907.4562 [hep-lat]];

\bibitem{Deuzeman:2008sc}
  A.~Deuzeman, M.~P.~Lombardo and E.~Pallante,
  {\em The physics of eight flavours},
  Phys.\ Lett.\  B {\bf 670}, 41 (2008)
  [arXiv:0804.2905 [hep-lat]];

\bibitem{Deuzeman:2009mh}
  A.~Deuzeman, M.~P.~Lombardo and E.~Pallante,
  {\em Evidence for a conformal phase in SU(N) gauge theories,}
  Phys.\ Rev.\  D {\bf 82}, 074503 (2010)
  [arXiv:0904.4662 [hep-ph]];

\bibitem{Itou:2010we}
  E.~Itou {\it et al.},
  {\em Search for the IR fixed point in the Twisted Polyakov Loop scheme (II)}
  arXiv:1011.0516 [hep-lat].

\bibitem{Jin:2010vm}
  X.~Y.~Jin and R.~D.~Mawhinney,
  {\em Evidence for a First Order, Finite Temperature Phase Transition in 8 Flavor
    QCD,}
  PoS {\bf LATTICE2010}, 055 (2010)
  [arXiv:1011.1511 [hep-lat]].

\bibitem{Hayakawa:2010yn}
  M.~Hayakawa, K.~I.~Ishikawa, Y.~Osaki, S.~Takeda, S.~Uno and N.~Yamada,
  {\em Running coupling constant of ten-flavor QCD with the Schr\'odinger
    functional method},
  arXiv:1011.2577 [hep-lat].

\bibitem{Hasenfratz:2010fi}
  A.~Hasenfratz,
  {\em Conformal or Walking? Monte Carlo renormalization group studies of SU(3)
    gauge models with fundamental fermions,}
  Phys.\ Rev.\  D {\bf 82}, 014506 (2010)
  [arXiv:1004.1004 [hep-lat]].

\bibitem{Hasenfratz:2009ea}
  A.~Hasenfratz,
  {\em Investigating the critical properties of beyond-QCD theories using Monte
    Carlo Renormalization Group matching,}
  Phys.\ Rev.\  D {\bf 80}, 034505 (2009)
  [arXiv:0907.0919 [hep-lat]].



\bibitem{Shamir:2008pb}
  Y.~Shamir, B.~Svetitsky and T.~DeGrand,
  {\em{Zero of the discrete beta function in SU(3) lattice gauge theory with color
  sextet fermions}},
  Phys.\ Rev.\  D {\bf 78}, 031502 (2008)
  [arXiv:0803.1707 [hep-lat]];
%
\bibitem{DeGrand:2008kx}
  T.~DeGrand, Y.~Shamir and B.~Svetitsky,
  {\em Phase structure of SU(3) gauge theory with two flavors of
    symmetric-representation fermions,}
  Phys.\ Rev.\  D {\bf 79}, 034501 (2009)
  [arXiv:0812.1427 [hep-lat]];
%
\bibitem{DeGrand:2010na}
  T.~DeGrand, Y.~Shamir and B.~Svetitsky,
  {\em Running coupling and mass anomalous dimension of SU(3) gauge theory with
    two flavors of symmetric-representation fermions,}
  Phys.\ Rev.\  D {\bf 82}, 054503 (2010)
  [arXiv:1006.0707 [hep-lat]].

\bibitem{Fodor:2009ar}
  Z.~Fodor, K.~Holland, J.~Kuti, D.~Nogradi and C.~Schroeder,
  {\em Chiral properties of SU(3) sextet fermions,}
  JHEP {\bf 0911}, 103 (2009)
  [arXiv:0908.2466 [hep-lat]].

\bibitem{Kogut:2010cz}
  J.~B.~Kogut and D.~K.~Sinclair,
  {\em Thermodynamics of lattice QCD with 2 flavours of colour-sextet quarks: A
    model of walking/conformal Technicolor,}
  Phys.\ Rev.\  D {\bf 81}, 114507 (2010)
  [arXiv:1002.2988 [hep-lat]].



\bibitem{Luscher:1996vw}
  M.~Luscher and P.~Weisz,
  {\em O(a) improvement of the axial current in lattice QCD
    to one-loop order  of perturbation theory,}
  Nucl.\ Phys.\  B {\bf 479} (1996) 429
  [arXiv:hep-lat/9606016].

\bibitem{Luscher:1996ug}
  M.~Luscher, S.~Sint, R.~Sommer, P.~Weisz and U.~Wolff,
  {\em Non-perturbative O(a) improvement of lattice QCD,}
  Nucl.\ Phys.\  B {\bf 491}, 323 (1997)
  [arXiv:hep-lat/9609035].

\bibitem{Karavirta:2010ym}
  T.~Karavirta, K.~Tuominen, A.~Mykkanen, J.~Rantaharju and K.~Rummukainen,
  {\em Non-perturbatively improved clover action for SU(2) gauge + fundamental and
    adjoint representation fermions,}
  PoS {\bf LATTICE2010} (2010) 064
  [arXiv:1011.1781 [hep-lat]].

\bibitem{Karavirta:2011mv}
  T.~Karavirta, A.~Mykkanen, J.~Rantaharju, K.~Rummukainen, K.~Tuominen,
  {\em Nonperturbative improvement of SU(2) lattice gauge theory with adjoint or fundamental flavors,}
  JHEP {\bf 1106 } (2011)  061.
  [arXiv:1101.0154 [hep-lat]].

\bibitem{Karavirta:2010ef}
  T.~Karavirta, K.~Tuominen, A.~-M.~Mykkanen, J.~Rantaharju, K.~Rummukainen,
  {\em Perturbative improvement of SU(2) gauge theory with two Wilson fermions in the adjoint representation,}
  PoS {\bf LATTICE2010 } (2010)  056.
  [arXiv:1011.2057 [hep-lat]].

\bibitem{Luscher:1992an}
  M.~Luscher, R.~Narayanan, P.~Weisz and U.~Wolff,
  {\em The Schrodinger functional: A Renormalizable probe for nonAbelian gauge
    theories,}
  Nucl.\ Phys.\  B {\bf 384}, 168 (1992)
  [arXiv:hep-lat/9207009].

\bibitem{Luscher:1992ny}
  M.~Luscher, R.~Narayanan, R.~Sommer, U.~Wolff, P.~Weisz,
  {\em Determination of the running coupling in the SU(2) Yang-Mills theory from first principles,}
  Nucl.\ Phys.\ Proc.\ Suppl.\  {\bf 30 } (1993)  139-148.

\bibitem{Luscher:1993gh}
  M.~Luscher, R.~Sommer, P.~Weisz {\it et al.},
  {\em A Precise determination of the running coupling in the SU(3) Yang-Mills theory,}
  Nucl.\ Phys.\  {\bf B413 } (1994)  481-502.
  [hep-lat/9309005].

  
\bibitem{DellaMorte:2004bc}
  M.~Della Morte, R.~Frezzotti, J.~Heitger, J.~Rolf, R.~Sommer and U.~Wolff
                  [ALPHA Collaboration],
  {\em Computation of the strong coupling in QCD with two dynamical flavours,}
  Nucl.\ Phys.\  B {\bf 713}, 378 (2005)
  [arXiv:hep-lat/0411025].

\bibitem{DellaMorte:2005kg}
  M.~Della Morte, R.~Hoffmann, F.~Knechtli, J.~Rolf, R.~Sommer, I.~Wetzorke and U.~Wolff                  [ALPHA Collaboration],
  {\em Non-perturbative quark mass renormalization in two-flavor QCD,}
  Nucl.\ Phys.\  B {\bf 729}, 117 (2005)
  [arXiv:hep-lat/0507035].

\bibitem{Capitani:1998mq}
  S.~Capitani, M.~Luscher, R.~Sommer and H.~Wittig  [ALPHA Collaboration],
  {\em Nonperturbative quark mass renormalization in quenched lattice QCD,}
  Nucl.\ Phys.\  B {\bf 544}, 669 (1999)
  [arXiv:hep-lat/9810063].

\bibitem{Sint:1995ch}
  S.~Sint and R.~Sommer,
  {\em The Running coupling from the QCD Schrodinger functional: A One loop
    analysis,}
  Nucl.\ Phys.\  B {\bf 465} (1996) 71
  [arXiv:hep-lat/9508012].

\bibitem{Sommer:1997jg}
  R.~Sommer,
  Nucl.\ Phys.\ Proc.\ Suppl.\  {\bf 60A}, 279 (1998)
  [arXiv:hep-lat/9705026].


\bibitem{Omelyan}
  I.P. Omelyan, I.M. Mryglod and R. Folk,
  {\em Symplectic analytically integrable decomposition algorithms: classification, derivation, and application to molecular dynamics, quantum and celestial mechanics simulations},
  Computer Physics Communications, Volume 151, Issue 3, 1 April 2003.

\bibitem{Takaishi:2005tz}
  T.~Takaishi and P.~de Forcrand,
  {\em Testing and tuning new symplectic integrators for hybrid Monte Carlo
    algorithm in lattice QCD}
  Phys.\ Rev.\  E {\bf 73} (2006) 036706
  [arXiv:hep-lat/0505020].

\bibitem{Brower:1995vx}
  R.~C.~Brower, T.~Ivanenko, A.~R.~Levi, K.~N.~Orginos,
  {\em Chronological inversion method for the Dirac matrix in hybrid Monte Carlo,}
  Nucl.\ Phys.\  {\bf B484 } (1997)  353-374.
  [hep-lat/9509012].


\bibitem{MILC}
  http://physics.utah.edu/$\sim$detar/milc.html 



\end{thebibliography}
\end{document}